\documentclass[pra,twocolumn,showpacs]{revtex4}
\usepackage{graphicx}
\usepackage{amsmath}

\newcommand{\E}{\operatorname{E}}

\begin{document}
\title{Single-Photon Switch Based on Rydberg Blockade}
\author{Simon Baur}
\author{Daniel Tiarks}
\author{Gerhard Rempe}
\author{Stephan D\"{u}rr}
\affiliation{Max-Planck-Institut f\"{u}r Quantenoptik, Hans-Kopfermann-Stra{\ss}e 1, 85748 Garching, Germany}


\begin{abstract}
All-optical switching is a technique in which a gate light pulse changes the transmission of a target light pulse without the detour via electronic signal processing. We take this to the quantum regime, where the incoming gate light pulse contains only one photon on average. The gate pulse is stored as a Rydberg excitation in an ultracold atomic gas using electromagnetically induced transparency. Rydberg blockade suppresses the transmission of the subsequent target pulse. Finally, the stored gate photon can be retrieved. A retrieved photon heralds successful storage. The corresponding postselected subensemble shows an extinction of 0.05. The single-photon switch offers many interesting perspectives ranging from quantum communication to quantum information processing.
\end{abstract}

\maketitle

The switch is the device that lies at the heart of digital signal processing which has revolutionized the fields of communication and computation. In both fields, optical techniques are increasingly gaining importance. For example, present-day high-bandwidth internet connections operate optically. In the field of computing, perspectives for optical techniques are being studied, too \cite{Miller:10, Caulfield:10}. They rely on all-optical switching and promise high bandwidth and low dissipated power. This creates a generic interest in the fundamental low-power limit of an all-optical switch, which is reached when the incoming gate pulse contains only one photon. Such a single-photon switch operates on the level of a single quantum and is hence well-suited for applications in quantum technology. For example, it offers perspectives for heralded quantum memories which will be essential for realizing quantum repeaters \cite{briegel:98}, for efficiently detecting optical photons in a nondestructive measurement \cite{Braginsky:96}, for generating Schr\"{o}dinger-cat states \cite{Gheri:97}, and for various other applications in the fields of quantum communication and quantum information processing \cite{nielsen:00, Bermel:06, Chang:07}.

The field of all-optical switching with a huge number of photons per gate pulse had traditionally been dominated by nonlinear optics with techniques such as saturable absorbers and optical bistability. Building a single-photon switch with those techniques would be very difficult because the nonlinearities in nonlinear crystals are tiny at the single-photon level. Lately, however, electromagnetically induced transparency (EIT) \cite{fleischhauer:05} enriched the field of all-optical switching.
If EIT is combined with Rydberg states \cite{Mohapatra:07}, one can use Rydberg blockade \cite{Jaksch:00, Tong:04} to create very large nonlinearities \cite{Pritchard:10, Dudin:12, Peyronel:12, Parigi:12, Maxwell:13, Hofmann:13}. This triggered a proposal for building single-photon quantum devices \cite{Gorshkov:11}. Experiments observed all-optical switching in different systems, see e.g.\ Refs.\ \cite{Dawes:05, Hwang:09, Bajcsy:09, Vo:12, Englund:12, Bose:12, Volz:12, Loo:12}. However, all these experiments required $\sim$20 or more incoming photons per gate pulse to obtain a clearly-visible switching effect. A very recent experiment demonstrated all-optical switching with 2.5 to 5 incoming photons based on normal-mode splitting in a cavity \cite{Chen:13}.

Here we experimentally demonstrate all-optical switching with a gate pulse that contains only one incoming photon on average, or even fewer. This gate pulse reduces the transmission of a subsequent target pulse by a factor of $\epsilon= 0.812\pm0.001$. To achieve this goal, we send the gate pulse into an ultracold atomic gas and store it as a Rydberg excitation using a slow-light technique based on Rydberg EIT. Next, the target pulse is sent through the atomic medium. Without the gate pulse, Rydberg EIT would result in high transmission of the target pulse. With the gate pulse, however, Rydberg blockade suppresses the transmission of the target pulse. After application of the target pulse, we can retrieve the stored gate excitation. This shows that coherence in the stored excitation survives the target pulse. Using the retrieval as a herald to indicate successful storage events, we obtain an extinction of $\epsilon= 0.051\pm0.004$ in the postselected subensemble. We study the dependence of $\epsilon$ on the numbers of incoming gate and target photons. The Rydberg blockade displays a lifetime of $\sim$60 $\mu$s if the target pulse is delayed relative to the gate pulse. The dephasing rate that limits the number of retrieved excitations depends linearly on the atomic density.

\begin{figure*}[!t]
\centering
\includegraphics[scale=1]{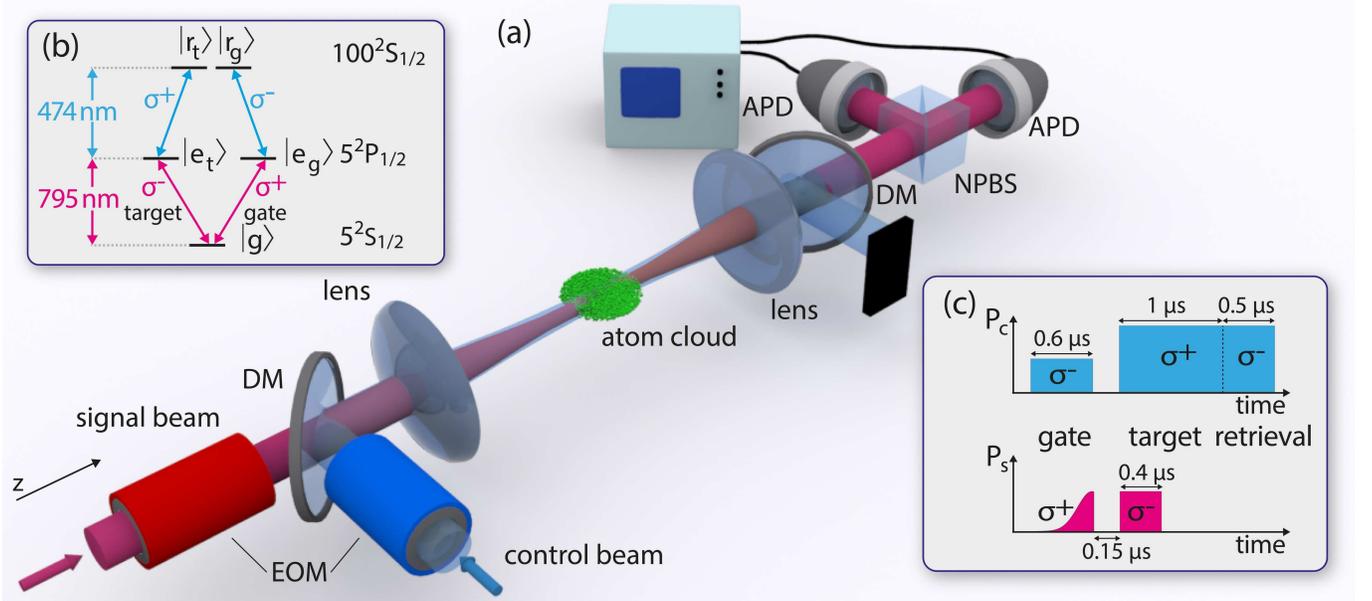}
\caption{\label{fig-scheme}
(a) Simplified scheme of the optical beam path. Signal and control beams for Rydberg EIT copropagate along the $z$ axis. After propagation through an ultracold gas of $^{87}$Rb atoms, a dichroic mirror (DM) splits off the control light, sending it onto a beam dump. The signal light is divided by a non-polarizing 50:50 beam splitter (NPBS) and detected on two avalanche photodiodes (APDs). Electro-optic modulators (EOMs) are used to set the incoming polarizations to either $\sigma^+$ or $\sigma^-$.
(b) Atomic level scheme. Signal light with polarizations $\sigma^+$ and $\sigma^-$ couples the ground state $|g\rangle= |5^2S_{1/2},$ $ F=1,$ $m_F=-1\rangle$ with the excited states $|e_g\rangle= |5^2P_{1/2}, F=2, m_F=0\rangle$ and $|e_t\rangle= |5^2P_{1/2}, F=2, m_F=-2\rangle$ for the gate and target pulse, respectively. Control light with polarizations $\sigma^-$ and $\sigma^+$ couples states $|e_g\rangle$ and $|e_t\rangle$ with Rydberg states $|r_g\rangle= |100^2S_{1/2}, m_J= m_I=-1/2\rangle$ and $|r_t\rangle= |100^2S_{1/2}, m_J=1/2, m_I=-3/2\rangle$ for the gate and target pulse, respectively.
(c) Timing of incoming light, see text.
}
\end{figure*}

Schemes of the experimental setup and the atomic levels are shown in Figs.\ \ref{fig-scheme}(a) and (b). Signal and control light have wavelengths of $\lambda_s=795$ nm and $\lambda_c=474$ nm and waists ($1/e^2$ radii of intensity) of $w_s=8$ $\mu$m and $w_c=12$ $\mu$m. The power of the control light is $P_c=32$ mW for target and retrieval and half as large for the gate. The ultracold gas consists of $N=2.2\times10^5$ atoms at a temperature of $T=0.43$ $\mu$K, which is a factor of $\sim$3 above the critical temperature for Bose-Einstein condensation. The atoms are held in a crossed-beam optical dipole trap at a wavelength of 1064 nm with measured trap frequencies of $(\omega_x,\omega_y,\omega_z)/2\pi= (136, 37, 37)$ Hz. All atoms are prepared in state $|g\rangle$. A magnetic field of $\sim$0.2 Gauss along the $z$ axis preserves the spin orientation. The efficiency for collecting and detecting a transmitted signal photon is 27\%. See Ref.\ \cite{supplement} for further details.

Fig.\ \ref{fig-scheme}(c) shows the timing sequence of the incoming light pulses. The gate pulse is followed by a dark time $t_d=0.15$ $\mu$s and then by the target pulse. Both pulses consist of light at the signal and control wavelengths. The signal light is resonant with an atomic transition, causing absorption. The control light creates EIT, thus suppressing the absorption of the signal light. The gate control light is switched off while a large part of the gate signal light is inside the medium due to a small group velocity. This stores gate signal photons in the medium in the form of Rydberg excitations.

To prevent the target control light from reading out these stored Rydberg excitations, the polarization of the control light is switched from $\sigma^-$ for the gate pulse to $\sigma^+$ for the target pulse. Hence, the target control light cannot couple the stored Rydberg excitations to any state in the $5^2P_{1/2}$ manyfold, because such a state would require $m_J=-3/2$, contradicting $J=1/2$. The signal light polarization is also switched. See Ref.\ \cite{supplement} for further details.

The long-range character of the van-der-Waals potential $V(r)= -C_6/r^6$ between Rydberg atoms causes Rydberg blockade. Here $r$ is the interatomic distance and $C_6$ is the van-der-Waals coefficient. Due to $V(r)$, the presence of a Rydberg excitation shifts the resonance frequency of the EIT feature for other incoming photons. This yields a blockade radius \cite{supplement} of $r_b= 14$ $\mu$m. For $r<r_b$, the resonance shift is larger than the width of the EIT feature and the system is shifted out of the EIT resonance, resulting in absorption. Our experiment is carried out in the regime $w_s\lesssim r_b$, where the blockade sphere surrounding a single Rydberg atom extends over the full transverse profile of the signal beam. Ideally, one would expect that a single Rydberg excitation stored during the gate pulse should reduce the transmission of the target signal beam to near zero. This brings us into a new regime in which we study the absorption that a propagating excitation experiences due to a stationary excitation stored during a previous pulse.

After the target signal pulse has left the medium, we switch the polarization of the control light back from $\sigma^+$ to $\sigma^-$. This retrieves the excitations stored during the gate pulse. We can use postselection conditioned on the detection of a retrieved photon as a powerful tool for exploring the full potential of Rydberg blockade as a mechanism for all-optical switching, eliminating the reduction of performance due to imperfect storage.

This gate-target pulse sequence is repeated with a cycle repetition time \cite{supplement} of $t_{\mathrm{cyc}}=100$ $\mu$s. Over the course of several thousand gate-target cycles, the atom number drops, so that a new atomic sample must be loaded.

\begin{figure}[!t]
\centering
\includegraphics[scale=1]{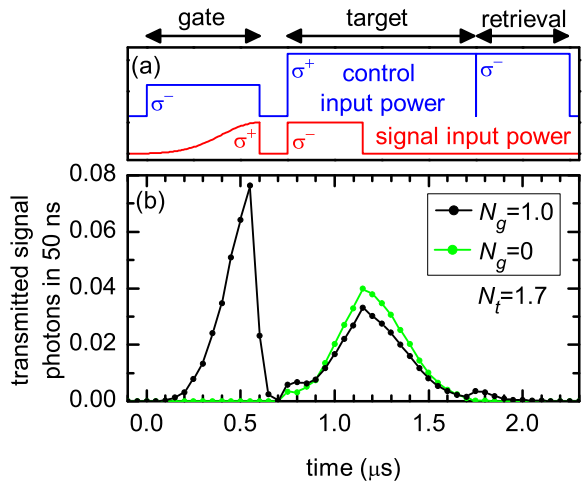}
\caption{\label{fig-time-traces}
(a) Input-power timing sequence.
(b) Single-photon switch. Black data show the average number of transmitted signal photons for an average number of incoming signal photons during the gate pulse of $N_g=1.0$. Green data show a reference with $N_g=0$. The extinction between black and green target-pulse data is $\epsilon= 0.812\pm0.001$. The deviation from $\epsilon=1$ is clearly observed, thus demonstrating a single-photon switch. The average number of incoming target signal photons is $N_t=1.7$. The subensemble postselected on the detection of a retrieved photon yields $\epsilon= 0.051\pm0.004$.
}
\end{figure}

Experimental results are shown in Fig.\ \ref{fig-time-traces}. The black data show two large peaks. The first peak shows an undesired nonzero transmission during the gate pulse due to imperfect storage. The second large peak shows the number of transmitted target photons which is reduced compared to the green reference data. There are also two smaller peaks in the black data: one at the beginning of the target interval, showing undesired partial readout of the stored gate excitation, the other at the beginning of the retrieval interval, showing the desired retrieval signal used for postselection.

To quantify how well the gate pulse reduces the transmission of target signal photons, we use the extinction
\begin{equation}
\epsilon
= \frac{N_{\text{trans}} \mbox{ with gate signal pulse}}{N_{\text{trans}} \mbox{ without gate signal pulse}}
,\end{equation}
where $N_{\text{trans}}$ denotes the mean number of transmitted target signal photons in one gate-target cycle. A reduction of $\epsilon$ below 1 is clearly observed in Fig.\ \ref{fig-time-traces}, thus realizing an all-optical switch. As the average number of signal photons in the incoming gate pulse $N_g$ is only 1.0, this measurement demonstrates a single-photon switch. The data in Fig.\ \ref{fig-time-traces} were averaged over $\sim 8\times10^6$ gate-target cycles \cite{supplement}.

As the signal light is derived from an attenuated laser beam, the incoming gate photons have a Poissonian number distribution so that there is a noticeable probability that more than one photon enters the medium. However, the probability for storing more than one photon is negligible due to Rydberg blockade among the gate photons before storage, as experimentally confirmed by measuring the pair-correlation function in a retrieval experiment \cite{supplement}.

If the storage of different gate photons is uncorrelated, then the number of excitations stored will be Poissonian, too. This is expected for small $N_g$ where blockade among gate photons has little relevance. Hence, the probability of storing zero excitations is $p_{s,0}= \exp(-\beta N_g)$ for small $N_g$. Here $\beta$ is the storage efficiency in the absence of Rydberg blockade. Obviously, $p_{s,0}$ sets a lower bound on the extinction $p_{s,0}\leq \epsilon$.

\begin{figure}[!t]
\centering
\includegraphics[scale=1]{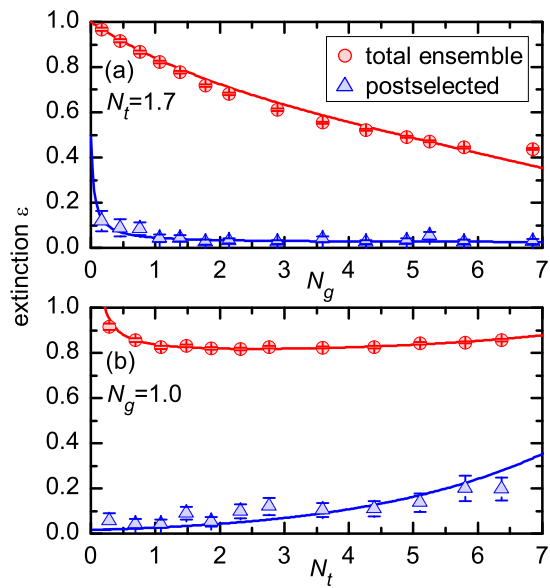}
\caption{\label{fig-Ng-and-Nt}
(color online)
(a) Dependence of the extinction $\epsilon$ on the incoming average photon number in the gate pulse $N_g$. Large $N_g$ reduces the probability of storing zero Rydberg excitations, resulting in improved average extinction in the total ensemble. The subensemble that is postselected conditioned on detecting a retrieved gate excitation shows a drastically improved extinction. This proves that the nonideal extinction in the total ensemble is dominantly limited by the storage efficiency.
(b) Dependence of the extinction $\epsilon$ on the incoming average photon number in the target pulse $N_t$. $\epsilon$ is fairly robust against changing $N_t$. All lines show fits to models from Ref.\ \cite{supplement}.
}
\end{figure}

Fig.\ \ref{fig-Ng-and-Nt}(a) shows an experimental study of $\epsilon(N_g)$. Note that even for $N_g=0.17$, we observe a deviation of $\epsilon$ from 1 by 4.5 standard errors in the total ensemble and by 20 standard errors after postselection. As the gate photons create blockade for each other, the simple estimate above is only applicable for small $N_g$. Hence, $\beta$ can be obtained from the absolute value of the slope $\beta=|d\epsilon/dN_g|$ at $N_g\to0$. The lines are fits of models of Ref.\ \cite{supplement}. The best-fit value is $\beta= 0.19$. The postselected subensemble shows a drastically improved extinction. For very small $N_g$, the postselected extinction deteriorates slightly. This is because for small $N_g$, the heralding probability decreases \cite{supplement} so that background counts during the retrieval interval contribute an increasing fraction to the heralded events.

Fig.\ \ref{fig-Ng-and-Nt}(b) shows the dependence of $\epsilon$ on $N_t$ at fixed target pulse duration. The dependence is rather weak, showing that the single-photon switch is fairly robust. The lines show fits to models from Ref.\ \cite{supplement}. The slight deterioration of $\epsilon$ for larger $N_t$ is due to the fact that scattering target signal photons reduces the atomic density, thus reducing the absorption. But this occurs only when averaging over large number of cycles for each atomic gas. The deterioration of $\epsilon$ for small $N_t$ is due to background photo detection events due to undesired readout of stored gate photons during the target pulse.

All above measurements were performed with a dark time between gate and target pulse of $t_d=0.15$ $\mu$s. If $t_d$ is increased, then the extinction $\epsilon$ decays with a $1/e$ time of 60 $\mu$s \cite{supplement}, again showing that this all-optical switch is fairly robust.

\begin{figure}[!t]
\centering
\includegraphics[scale=1]{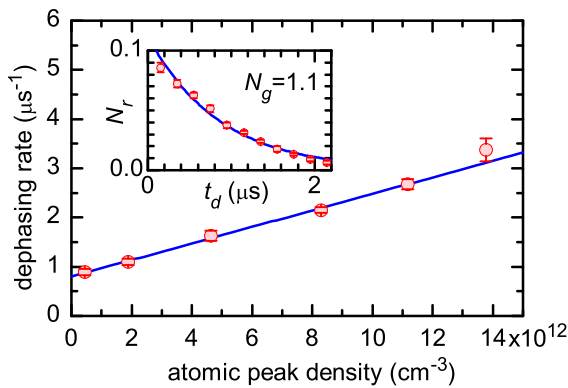}
\caption{\label{fig-dephasing-rate}
(color online)
The measured dephasing rate increases linearly with atomic density. Inset: Measurement of the dephasing rate at a peak density of $\sim 2\times10^{12}$ cm$^{-3}$, where the experiment is normally operated. The retrieved photon number $N_r$ decays as a function of the dark time $t_d$ between gate and retrieval pulse in the absence of a target pulse. An exponential fit yields the $1/e$ dephasing rate.
}
\end{figure}

For comparison, note that the retrieved signal in the absence of a target pulse decays with a $1/e$ time of $\sim0.9$ $\mu$s. This much shorter time scale is because retrieval is based on phase-coherent collective directed emission, whereas the blockade merely needs Rydberg population. The ability to perform postselection crucially relies on a sufficiently long coherence time. We find that the coherence time depends linearly on the density of surrounding ground-state atoms, as shown in Fig.\ \ref{fig-dephasing-rate}. We attribute this to a shift of the EIT resonance due to collisions between a Rydberg atom and surrounding ground-state atoms. Such a shift of $\sim10$ MHz at a density of $\sim 10^{14}$ cm$^{-3}$ was observed in Ref.\ \cite{Balewski:13}. The inhomogeneity of the atomic sample converts this shift into a dephasing process. At zero density, the line extrapolates to a dephasing rate of 0.8 $\mu$s$^{-1}$. Due to thermal motion of the atoms we expect 0.14 $\mu$s$^{-1}$, indicating that further decoherence mechanisms play a role. If future work can identify these mechanisms and remove them, we expect improvements in storage efficiency, heralding probability, and EIT transmission. This offers room for substantial improvements of the overall performance of the all-optical switch.

The work presented here opens the door to the new world of single-photon switching. With better performance, this will bring exciting perspectives in quantum information processing into reach. First, heralding successful storage is interesting for quantum memories. Storage times could be improved by subsequent transfer of the population into long-lived ground states. Second, the presence or absence of one gate photon could be mapped to the absence or presence of many transmitted target photons, respectively. Discriminating between the latter cases is easy, even at low detector efficiency. This could allow for detection of an optical photon with high sensitivity. Third, if the stored photon is eventually retrieved, then the detection of many target photons will represent a nondestructive detection of a single optical photon \cite{Reiserer:13}. Fourth, if the incoming gate pulse contains a coherent superposition of zero and one photon, then the single-photon switch can create a Schr\"{o}dinger-cat type coherent superposition of states with macroscopically different target photons numbers. Fifth, a photonic quantum-logic gate could be built based on this single-photon switch. For applications four and five, dissipation and decoherence must be kept low, which at first glance seems to contradict switching between transmission and absorption. However, if our switch is placed inside an optical resonator, resonant with the signal light, then transmission inside the atomic gas will lead to transmission through the resonator, whereas absorption inside the atomic gas will lead to reflection from the first mirror. This will convert the transmission-absorption switch into a transmission-reflection switch which could operate at low dissipation and decoherence.

We thank D. Fauser for assistance during an early stage of the experiment. We acknowledge fruitful discussions on dephasing with T. Pfau. This work was supported by the DFG via NIM and via SFB 631.

\clearpage
\renewcommand{\theequation}{S\arabic{equation}}
\renewcommand{\thefigure}{S\arabic{figure}}
\setcounter{equation}{0}
\setcounter{figure}{0}

\section*{APPENDIX}

This appendix discusses the properties of the atomic gas in Sec.\ \ref{sec-atoms},  EIT in Sec.\ \ref{sec-EIT}, various aspects of Rydberg blockade in Sec.\ \ref{sec-blockade}, the heralding probability in Sec.\ \ref{sec-herald}, the choice of atomic transitions in Sec.\ \ref{sec-transitions}, and the gate-target cycles in Sec.\ \ref{sec-gate-target-cycles}. We discuss simple models, which capture the dominant physical effects, explain all experimentally observed features, and have the value of being intuitively accessible. More detailed modeling lies beyond the scope of this work.

\section{Properties of the Atomic Gas}

\label{sec-atoms}

In the absence of control light, the atomic cloud with atom number $N=2.2\times10^5$ and temperature $T=0.43$ $\mu$K held in a trap with frequencies $(\omega_x,\omega_y,\omega_z)/2\pi= (136, 37, 37)$ Hz is estimated to have root-mean-square (rms) radii $(\sigma_x,\sigma_y,\sigma_z)= (7.5,28,28)$ $\mu$m and a peak density of $\varrho_p= 2.4\times10^{12}$ cm$^{-3}$. Averaging the transmission of signal light over the transverse profile of the signal beam yields $\langle T\rangle$. We define the effective optical depth by setting $\langle T\rangle= e^{-OD}$. This yields the estimates $OD_g= 3.5$ for $\sigma^+$ polarized signal light (gate pulse) and $OD_t= 10$ for $\sigma^-$ polarized signal light (target pulse). The absorption cross section for signal light on the target transition is a factor of 6 larger than for the gate transition. In a homogeneous medium, $OD_t/OD_g$ would also equal 6, but transverse averaging changes this ratio.

The atomic ground state $|g\rangle$ has a dynamic polarizability of $\alpha= -163\times4\pi\epsilon_0a_0^3$ \cite{Saffman:10} for control light at $\lambda_c=473.9$ nm, where $\epsilon_0$ is the vacuum permittivity and $a_0$ the Bohr radius. The resulting optical dipole potential is repulsive so that it pushes atoms away from the trap center. After time averaging over one optical period, the spatial maximum of the potential is estimated to be $V_0= -\alpha E_{c,0}^2/4$, where $E_{c,0}$ is the electric field amplitude of the control beam. At a control power of $P_c=32$ mW this yields $V_0= k_B\times5.1$ $\mu$K, where $k_B$ is the Boltzmann constant. Within the cycle repetition time of $t_{\text{cyc}}=100$ $\mu$s, control light is turned on for 1.5 $\mu$s with $P_c=32$ mW and for 0.6 $\mu$s with $P_c=16$ mW. $t_{\text{cyc}}$ is much shorter than all trap oscillation periods and the distance traveled by an atom within a few microseconds is negligible, so that it is justified to consider the potential only after time averaging over one gate-target cycle. Its spatial maximum is $\langle V_0\rangle= 0.018V_0= k_B\times0.09$ $\mu$K. This is small compared to the temperature of the atoms. The repulsion is estimated to reduce the density at the trap center as well as $OD_g$ by $\sim$10\%. This is so small that we neglect the effect of the repulsive potential throughout our analysis.

After starting the gate-target cycles, the timescale for a new equilibrium of the atomic density to establish is expected to be on the order of the slowest trap oscillation period, which is $\sim$30 ms. This is confirmed by experimental observations. To reduce averaging over different transmissions, we process data only between 50 and 950 ms after starting the gate-target cycles for most of our measurements. For long times, the scattering of signal photons causes loss of atoms due to evaporation from the shallow dipole trap. The effect of this is reduced by terminating the gate-target cycles after 950 ms, unless otherwise stated.

\section{Electromagnetically Induced Transparency}

\label{sec-EIT}

To estimate the control-light Rabi frequency $\Omega_c$, we start from the radial integral $\langle r\rangle_{5p}^{ns}= 0.014\times (50/n)^{3/2} a_0$ for $^{87}$Rb \cite{Saffman:10}. Textbook angular momentum algebra \cite{edmonds:63} yields electric dipole matrix elements $d_g= e\langle r\rangle_{5p}^{100s} /3 = 1.6\times10^{-3} ea_0$ and $d_t= e \langle r\rangle_{5p}^{100s} \sqrt2/3 = 2.3\times10^{-3} ea_0$ for the control transitions during gate and target pulse, respectively. Here $e$ is the elementary charge. The Rabi frequency is $\Omega_c= d E_{c,0}/\hbar$. The waist $w_c=12$ $\mu$m and the powers $P_{c,g}=16$ mW and $P_{c,t}=32$ mW yield $E_{c,g,0}= 0.23$ MV/m and $E_{c,t,0}= 0.32$ MV/m as well as $\Omega_{c,g}/2\pi= 4.7$ MHz and $\Omega_{c,t}/2\pi= 9.4$ MHz for gate and target pulse, respectively.

Next, we estimate the full width at half maximum (FWHM) $\Delta_T$ of the spectral transmission window in EIT. Considering a homogeneous gas and ignoring dephasing, the usual model for EIT \cite{fleischhauer:05} yields $\Delta_T= \Delta\omega_{\text{trans}}\sqrt{\ln2}= \Omega_c^2\sqrt{\ln2}/\Gamma\sqrt{OD}$, where $\Gamma= 2\pi\times5.75$ MHz is the decay rate of state $|e_{g/t}\rangle$. The above estimates for $OD$ and $\Omega_c$ yield $\Delta_{T,g}= 2\pi\times 1.7$ MHz and $\Delta_{T,t}= 2\pi\times 4.0$ MHz.

In the absence of dephasing, the maximum transmission at the EIT resonance $T_0=e^{-OD_{\text{EIT}}}$ should ideally be unity. But with a dephasing rate of $\gamma_{21}=1.1$ $\mu$s$^{-1}$, see Fig.\ \ref{fig-dephasing-rate}, the model of Ref.\ \cite{fleischhauer:05} predicts $OD_{\text{EIT},g}= 0.8$ and $OD_{\text{EIT},t}= 0.7$ for gate and target light, respectively.

\begin{figure}[!t]
\includegraphics[scale=0.95]{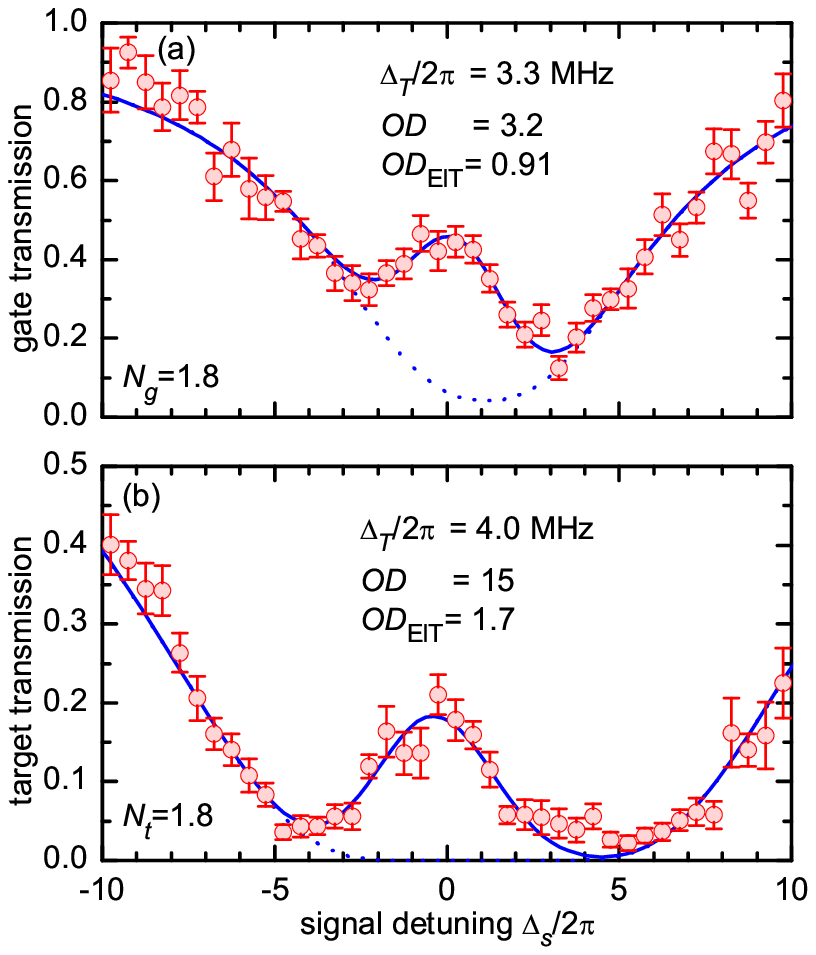}
\caption{\label{fig-sup-spectra}
\textbf{EIT spectra.} (a) Gate transition. The solid line shows a fit of Eq.\ \eqref{EIT-fit}.
The dotted line shows the same curve without the Gaussian peak.
(b) Same for the target transition. Polarizations of signal and control beams as well as control power are chosen as for the single-photon switch, i.e.\ they differ between (a) and (b). However, the rectangular signal pulse shape with duration 0.40 $\mu$s and the incoming signal photon number of 1.8 are identical in (a) and (b). Our choice of polarizations creates a much larger $OD$ for the target transition.
}
\end{figure}

Fig.\ \ref{fig-sup-spectra} shows measured EIT spectra for (a) the gate and (b) the target transition. Data were evaluated between 400 and 600 ms after starting the gate-target cycles. The solid line shows a fit of the simple, empiric model
\begin{multline}
\label{EIT-fit}
T=\exp\left(\frac{-OD}{1+(2(\Delta_s-\Delta_0)/\Gamma)^2}\right)
\\
+ T_0 \exp\left(-\frac{4(\Delta_s-\Delta_1)^2}{\Delta_T^2}\ln2\right)
\end{multline}
to the data. The first term is the atomic absorption line without EIT, where $OD$ is the optical depth and $\Delta_s= \omega_s-\omega_{s,\text{res}}$ is the detuning of the signal light. The second term empirically represents EIT as a Gaussian transmission feature. $\Delta_0$ and $\Delta_1$ determine the centers of the features. If $\omega_{s,\text{res}}$ were accurately calibrated and the control laser were exactly on resonance, then $\Delta_0$ and $\Delta_1$ would both vanish. The best-fit values with fixed $\Gamma$ are quoted in the figure. The agreement with the above estimates for $OD$, $\Delta_T$, and $OD_{\text{EIT}}$ is fair. The deviations are partly because Rydberg blockade causes a noticeable reduction of $T_0$ at $N_t=1.8$, see Sec.\ \ref{sec-propagation}. The remaining deviations are probably because the model ignores the inhomogeneity of the atomic density.

Frequency noise on the signal and control lasers makes a negligible contribution to the observed value of $\Delta_T/2\pi$, because the measured rms linewidths of the lasers of $\Delta\nu_s= 0.1$ MHz and $\Delta\nu_c= 0.2$ MHz are too small. The magnetic field of 0.2 G is estimated to cause a Zeeman energy splitting of $2\pi\hbar\times0.6$ MHz between the Rydberg states $|r_{g/t}\rangle$. This is also small compared to $\hbar\Delta_T$.

For later reference, we now briefly discuss absorption length and group velocity. The absorption length is $l_{a,t}= 1/\varrho\sigma_t$, where $\varrho$ is the atomic density, $\sigma_t= 3\xi_t\lambda_s^2/2\pi$ the photon absorption cross section on the target signal transition at $\lambda_s= 795$ nm, and $\xi_t= 1/2$ the branching ratio for decay of state $|e_t\rangle$ on the target signal transition. Approximating the medium as homogeneous with density $\varrho= \varrho_p/2$ with $\varrho_p= 2.4\times 10^{12}$ cm$^{-3}$ from Sec.\ \ref{sec-atoms}, we estimate $l_{a,t}= 5\ \mu\text{m}$. Based on $l_{a,t}$, the group velocity $v_g$ is estimated to be \cite{fleischhauer:05} $v_g\approx \Omega_{c,t}^2 l_{a,t}/\Gamma= 0.5 $ km/s when ignoring dephasing. This agrees fairly well with the target pulse delay of $t_{\text{delay},t}= 0.25$ $\mu$s observed in Fig.\ \ref{fig-time-traces}, which suggests a group velocity of $v_g= \sqrt{2\pi} \sigma_z/t_{\text{delay},t}= 0.3$ km/s.

\section{Rydberg Blockade}

\label{sec-blockade}

To calculate the blockade radius, we follow the definition $r_b= |2C_6\Gamma/\hbar\Omega_c^2|^{1/6}$ of Ref.\ \cite{Peyronel:12}, where $C_6=-3.9 \times 10^{23} E_h a_0^6$ \cite{Singer:05} is the van-der-Waals coefficient for the $100^2S_{1/2}$ state, $E_h=\hbar^2/m_ea_0^2$ the Hartree energy, and $m_e$ the electron rest mass. Combination with the above Rabi frequencies yields blockade radii of $r_{b,g}=18$ $\mu$m and $r_{b,t}=14$ $\mu$m for gate and target pulse, respectively. $r_{b,t}=14$ $\mu$m is the relevant number for the operation of our all optical switch, whereas $r_{b,g}=18$ $\mu$m describes self-blockade of the gate pulse.

We compare this result with two other typical length scales to understand in what regime the experiment is operated. Comparison with the absorption length $l_{a,t}= 5\ \mu\text{m}\lesssim 2r_b$ shows that a target photon subject to Rydberg blockade due to a stored gate excitation will experience substantial absorption within a distance of $2r_b$. Comparison with the signal beam waist $w_s=8\ \mu\text{m} \lesssim r_b$ shows that our experiment is in the one-dimensional regime, where the typical transverse distance between two excitations is insufficient to prevent blockade.

\subsection{Rydberg Blockade in Propagation}

\label{sec-propagation}

\begin{figure}[!t]
\includegraphics[scale=0.95]{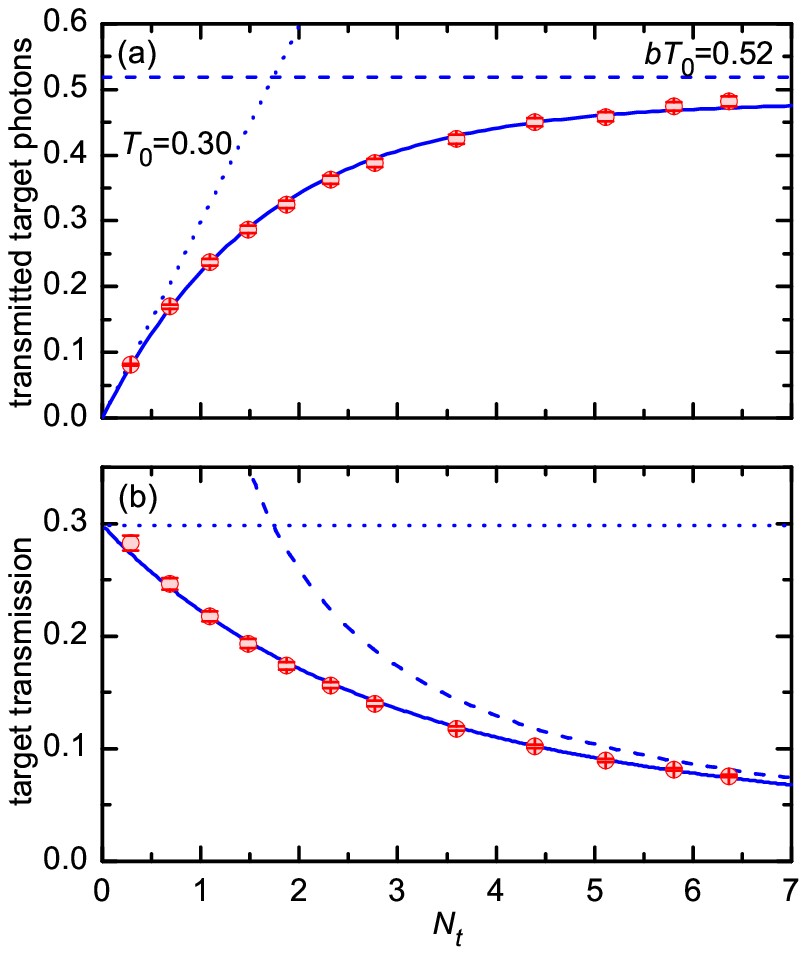}
\caption{\label{fig-sup-Nt}
\textbf{Rydberg blockade for copropagating excitations.} (a) Target photons create Rydberg blockade for each other for fixed target pulse duration and $N_g=0$. The solid line shows a fit of Eq.\ \eqref{propagation-N-out-rapid}. For small $N_t$, the number of transmitted target photons is linear in $N_t$ (dotted line). For large $N_t$, however, the number of transmitted photons levels off (approaching the dashed line) due to Rydberg blockade. (b) Same data in units of transmission.
}
\end{figure}

It is worth noting that Fig.\ \ref{fig-Ng-and-Nt}(b) shows only a moderate change in the extinction as a function of $N_t$, whereas the numerator and denominator in Eq.\ (1) individually each show a much stronger change. To clarify this point, Fig.\ \ref{fig-sup-Nt}(a) shows the number $N_{\text{trans}}$ of transmitted target signal photons as a function of the incoming photon number $N_t$ at the EIT resonance with $N_g=0$. For constant transmission, one would expect a linear behavior. This is observed only for small $N_t$. The fact that the transmitted photon number levels off for large $N_t$ is a signature of Rydberg blockade in EIT, similar to Refs.\ \cite{Pritchard:10, Peyronel:12}.

To model the propagation of excitations in the presence of Rydberg blockade, we consider a signal light pulse with constant input intensity and duration $t_p$ and divide it into $b\approx t_p/\tau_c$ bins of duration $\tau_c= 0.23$ $\mu$s which is the measured rms width of the antibunching feature, see Sec.\ \ref{sec-correlation}.

We assume a homogeneous medium with absorption coefficients $\alpha$ without EIT and $\alpha_1$ at the EIT resonance. The probability $p_n(z)$ of finding $n$ photons in a given bin after propagating a distance $z$ evolves according to
\begin{subequations}
\label{propagation-dpn-dz}
\begin{align}
\partial_z p_n
&= \alpha(-np_n+(n+1)p_{n+1})
,\qquad n\geq2
\\
\partial_z p_1
&= -\alpha_1 p_1+ 2 \alpha p_2
\\
\partial_z p_0
&= \alpha_1 p_1
.\end{align}
\end{subequations}
Here, we assumed that each photon is absorbed with coefficient $\alpha$ as long as the bin contains more than one photon, whereas a photon that is alone in the bin is absorbed with coefficient $\alpha_1$. We refer to this model as the \emph{binning approximation}. In addition to its one-dimensional character and the assumption of a homogeneous medium, the model also neglects crosstalk between different bins so that the size and the dispersion of longitudinal optical wave packets are not taken into account.

The incoming photon number in a given bin has a Poisson distribution with mean value
\begin{align}
\mu_0= \frac{N_{\text{in}}}{b}
.\end{align}
With Eq.\ \eqref{propagation-dpn-dz}, the probabilities evolve into
\begin{align}
p_n(z)= \frac{\mu^n(z)}{n!} e^{-\mu(z)}
,\qquad
n\geq2
\end{align}
with
\begin{align}
\mu(z)= \mu_0 e^{-\alpha z}
\end{align}
and into $p_1(z)= e^{-\alpha_1z}(p_1(0)+2\alpha\int_0^z p_2(\widetilde z)e^{\alpha_1\widetilde z}d\widetilde z)$. $p_0(z)= (1+\mu(z))e^{-\mu(z)}-p_1(z)$ follows from normalization. To calculate $p_1(z)$, we expand $e^{-\mu(\widetilde z)}$ into a power series $\sum_{k=0}^\infty (-\mu(\widetilde z))^k/k!$. The resulting integrals can be solved analytically, yielding $p_1(z)= e^{-\alpha_1z}(\mu_0e^{-\mu_0}+\alpha\sum_{k=0}^\infty (-\mu_0)^{k+2} (e^{(\alpha_1-(k+2)\alpha)z}-1)/k!(\alpha_1-(k+2)\alpha))$. In our experiment $\alpha_1/\alpha =OD_{\text{EIT}}/OD \ll1$. Hence, we neglect $\alpha_1$ in the denominator and obtain
\begin{align}
p_1(z)= e^{-\alpha_1z}(1-e^{-\mu_0})-1+(1+\mu(z))e^{-\mu(z)}
.\end{align}
The mean photon number in the bin is $N_{\text{bin}}(z)= \sum_{n=0}^\infty np_n(z)= p_1(z) + \mu(z)(1-e^{-\mu(z)})$. We obtain
\begin{align}
\label{propagation-N-bin}
N_{\text{bin}}(z) = e^{-\alpha_1z} (1-e^{-\mu_0})-1+\mu(z)+e^{-\mu(z)}
.\end{align}
After transmission through the medium of length $L$, the mean number of photons in the complete pulse is
\begin{align}
N_{\text{out}} = b N_{\text{bin}}(L)
.\end{align}
In Fig.\ \ref{fig-sup-Nt}, we have $N_g\leq7$, $b\sim1.7$, and $\alpha L = OD \geq 10$ so that $\mu(L)\leq 2\times10^{-4}\ll1$. Hence, we approximate $\mu(L)\sim0$ and obtain (see also Fig.\ 3(a) of Ref.\ \cite{Peyronel:12})
\begin{align}
\label{propagation-N-out-rapid}
N_{\text{out}} = bT_0(1-e^{-N_{\text{in}}/b})
,\end{align}
where $T_0 =e^{-\alpha_1L}= e^{-OD_{\text{EIT}}}$. A fit of Eq.\ \eqref{propagation-N-out-rapid} in Fig.\ \ref{fig-sup-Nt}(a) agrees well with the data and yields best-fit values $T_0 =0.30$ and $b=1.6$. The latter agrees well with the estimate $b\approx t_p/\tau_c= 0.4\ \mu\text{s}/0.23\ \mu\text{s}=1.7$.

A less rigorous argument can be used to derive Eq.\ \eqref{propagation-N-out-rapid} with less effort. To this end, note that $\mu(L)\ll1$ implies that only a small fraction of the distance $L$ is needed to achieve almost perfect Rydberg blockade. One can approximate this distance as infinitesimally short. We call this the \emph{rapid-blockade approximation}. As a result, linear absorption has no effect during this infinitesimal distance. Each bin with a nonzero initial photon number will contain one photon after this infinitesimal distance. The mean number of photons in one bin after this infinitesimal distance is $\sum_{n=1}^\infty p_n(0)= 1-e^{-\mu_0}$. For the whole pulse, the mean photon number surviving this infinitesimal distance is $b(1-e^{-\mu_0})$. This photon number is subject to linear absorption for the subsequent propagation through the medium, creating a factor $T_0$, thus yielding Eq.\ \eqref{propagation-N-out-rapid}.

\subsection{Rydberg Blockade in Storage}

\label{sec-storage}

We now extend the above model to include storage, in order to obtain a fit to Fig.\ \ref{fig-Ng-and-Nt}(a). Let $N_{\text{in}}$, $N_b$, and $N_s$ denote the mean numbers of incoming excitations, of excitations immediately before switching off the control power, and of stored excitations, respectively. $N_b$ is less than $N_{\text{in}}$ due to absorption during propagation inside the medium before switching off the control power. This absorption has a linear contribution and a contribution due to Rydberg blockade that the gate photons create for each other. However, we neglect blockade while switching off the control power because the switch off is fast. We define three storage efficiencies
\begin{align}
\eta_s(N_{\text{in}}) = \frac{N_s}{N_{\text{in}}}
,&&
\beta = \lim_{N_{\text{in}}\to0} \eta_s
,&&
\eta_{sb} = \frac{N_s}{N_b}
.\end{align}
Here, $\eta_s$ is the overall storage efficiency, $\beta$ the storage efficiency in the absence of Rydberg blockade, and $\eta_{sb}$ the storage efficiency in the absence of Rydberg blockade and linear absorption. Note that $\eta_s$ depends on $N_{\text{in}}$ due to blockade, whereas $\eta_{sb}$ is constant because we neglect blockade during the switch off.

We begin by calculating $N_b$ from $N_{\text{in}}$. It is obviously obtained by taking the sum over all bins $N_b= \sum N_{\text{bin}}(z)$ with $N_{\text{bin}}(z)$ from Eq.\ \eqref{propagation-N-bin}. For simplicity, we assume that the signal-light pulse is rectangular with spatial length $L_p= t_pv_g$, where $v_g$ is the group velocity. Hence, each bin has a spatial length $L_p/b$ and all bins have the same mean incoming photon number $\mu_0$. We assume that storage takes place when the end of the pulse has just entered the medium with length $L\geq L_p$. Assuming $b\gg1$, we approximate the sum over all bins by the integral $N_b=(b/L_p)\int_0^{L_p}N_{\text{bin}} dz$ and obtain
\begin{multline}
\label{storage-N-b}
N_b=
\frac{b}{L_p}\left( \frac{1-e^{-\alpha_1L_p}}{\alpha_1} (1-e^{-\mu_0})
\right. \\ \left.
-L_p+\frac{\mu_0-\mu(L_p)}{\alpha}+\frac{\E_1(\mu(L_p))-\E_1(\mu_0)}{\alpha}\right)
,\end{multline}
where $\E_1(z)= \int_z^\infty dt \, e^{-t}/t$ is the exponential integral.

After calculating $N_b$ from $N_{\text{in}}$, we now turn to the calculation of $N_s$. As stated above, we neglect blockade while switching off the control power. As a result, excitations present immediately before storage are stored independently of each other with a storage efficiency $\eta_{sb}= N_s/N_b$ that is independent of $N_b$. Hence, the probability of storing zero excitations is $p_{s,0}= \sum_{n=0}^\infty(1-\eta_{sb})^np_{b,n}$, where $p_{b,n}$ denotes the probability of having $n$ excitations in the total pulse immediately before switching off the control power. In our experiment $\eta_{sb}\ll1$. Neglecting terms of order $\mathcal O(\eta_{sb}^2)$, we obtain $p_{s,0}= 1-\eta_{sb} N_b$.

We now consider a subsequent target pulse. We assume that one stored gate excitation causes perfect Rydberg blockade for the target pulse so that $\epsilon= p_{s,0}$ and
\begin{align}
\label{storage-epsilon}
\epsilon(N_g)
= 1-\eta_{sb} N_b
.\end{align}
Combination of Eqs.\ \eqref{storage-N-b} and \eqref{storage-epsilon} yields an initial slope of
\begin{align}
\label{storage-beta}
\beta
= \lim_{N_{\text{in}}\to0} \left| \frac{\partial \epsilon}{\partial N_{\text{in}}} \right|
= \eta_{sb} \frac{1-e^{-\alpha_1L_p}}{\alpha_1L_p}
.\end{align}
The solid line in Fig.\ \ref{fig-Ng-and-Nt}(a) shows a fit of Eqs.\ \eqref{storage-N-b} and \eqref{storage-epsilon} to the data. In our experiment, we estimate that the pulse length inside the medium $L_p= v_gt_p$ is similar to the length of the medium $L\sim \sqrt{2\pi}\sigma_z$. After division by $v_g$, this is equivalent to the experimental observation that the pulse duration $t_p$ is similar to the pulse delay. Hence, we approximate $L_p=L$ and keep the parameters $OD_{\text{EIT}}= \alpha_1 L =0.91$ and $OD=\alpha L =3.2$ fixed at the values independently measured for the gate pulse in Sec.\ \ref{sec-EIT}. The resulting best-fit values are $b= 2.0$ and $\eta_{sb}= 0.29$, so that Eq.\ \eqref{storage-beta} yields $\beta= 0.19$.

To compare the best-fit value of $b$ to an independent estimate, we note that the gate signal pulse is shaped as a Gaussian that is cut off in the center. Without cutting, its rms width would be 0.2 $\mu$s. We approximate this as a rectangular pulse with duration $t_p=0.2$ $\mu$s. As a coarse approximation, we use the value of $\tau_c$ measured for the target light and obtain the estimate $b\approx t_p/\tau_c= 0.2\ \mu\text{s}/0.23\ \mu\text{s}= 0.9$. The agreement with the best-fit value is fair.

We now compare this model with the rapid-blockade approximation of Sec.\ \ref{sec-propagation} which converts Eq.\ \eqref{propagation-N-bin} into $N_{\text{bin}}(z) = e^{-\alpha_1z} (1-e^{-\mu_0})$ for all $z$ inside the medium. Spatial integration then yields a simple exponential model
\begin{align}
\label{storage-N-b-rapid}
N_b=
b \frac{1-e^{-\alpha_1L_p}}{\alpha_1 L_p} (1-e^{-\mu_0})
.\end{align}
Combined with Eq.\ \eqref{storage-epsilon}, this model also fits the data in Fig.\ \ref{fig-Ng-and-Nt}(a) well. The best-fit values are $b=3.2$ and $\eta_{sb}=0.31$, similar to the previous fit.

\subsection{Storage and Postselection}

The postselected data in Fig.\ \ref{fig-Ng-and-Nt}(a) are almost independent of $N_g$, showing that the nonzero probability of storing zero excitations is the dominant limiting factor for $\epsilon$ in the total ensemble.

The only visible trend in the postselected data is a slight deterioration of $\epsilon$ for small $N_g$. This is due to the fact that there are background counts during the retrieval interval. For vanishing $N_g$, the heralding probability $p_h(N_g)$ drops as discussed in Sec.\ \ref{sec-herald}, but it does not drop to exactly zero. Instead, we experimentally find $p_h(0)=1.4\times10^{-4}$. If such a background event incorrectly heralds a gate-target cycle, then $N_{\text{trans}}$ will simply be that of the total ensemble. This is easily modeled by
\begin{subequations}
\label{postselection-epsilon-post}
\begin{align}
\epsilon^{\text{post}}(N_g)
&= \frac{\left(1-q\right) N_{\text{trans,ideal}}^{\text{post}}+q N_{\text{trans}}^{\text{total}}(N_g)}
{N_{\text{trans}}^{\text{post}}(N_g=0)}
,\\
q(N_g)
&= \frac{p_h(0)}{p_h(N_g)}
.\end{align}
\end{subequations}
The denominator results immediately from the definition of $\epsilon$. The numerator is the weighted sum of two contributions: one contribution $N_{\text{trans,ideal}}^{\text{post}}$ that we would obtain for $p_h(0)=0$ and the other contribution $N_{\text{trans}}^{\text{total}}(N_g)$ that expresses the fact that if a background event incorrectly causes heralding of a gate-target cycle, then the transmission will be that of the total ensemble.

The two functions $p_h(N_g)$ and $N_{\text{trans}}^{\text{total}}(N_g)$ are known from Secs.\ \ref{sec-storage} and \ref{sec-herald}. Fitting functions to the data sets discussed in those sections, we obtain smooth fit curves for $p_h(N_g)$ and $N_{\text{trans}}^{\text{total}}(N_g)$. Keeping the parameters of those models fixed, we have only two parameters $\epsilon_{\text{ideal}}= N_{\text{trans,ideal}}^{\text{post}}/N_{\text{trans}}^{\text{post}}(N_g=0)$ and $N_{\text{trans}}^{\text{post}}(N_g=0)$ in Eq.\ \eqref{postselection-epsilon-post}. A fit of Eq.\ \eqref{postselection-epsilon-post} to the data in Fig.\ \ref{fig-Ng-and-Nt}(a) yields best-fit values $\epsilon_{\text{ideal}}=0.022\pm0.003$ and $N_{\text{trans}}^{\text{post}}(N_g=0)=0.7\pm0.2$. The model fits well to the data showing that the origin of the slight deterioration of the postselected $\epsilon$ for small $N_g$ is well understood. The value of $\epsilon_{\text{ideal}}$ also shows that the data with $N_g=1$ still suffer somewhat from this deterioration and that the true performance of the single-photon switch is better than the directly measured value of $\epsilon$ at $N_g=1$.

\subsection{Rydberg Blockade in the All-Optical Switch}

\label{sec-switch}

To model the data in Fig.\ \ref{fig-Ng-and-Nt}(b), we first consider the postselected subensemble. The transmitted photon number for the postselected subensemble can be written as
\begin{align}
\label{switch-OD-b-def}
N_{\text{trans}}= N_t e^{-OD_b}
,\end{align}
where $OD_b$ is the optical depth that a propagating target excitation experiences in the presence of one stored gate excitation.

\begin{figure}[!t]
\includegraphics[scale=0.95]{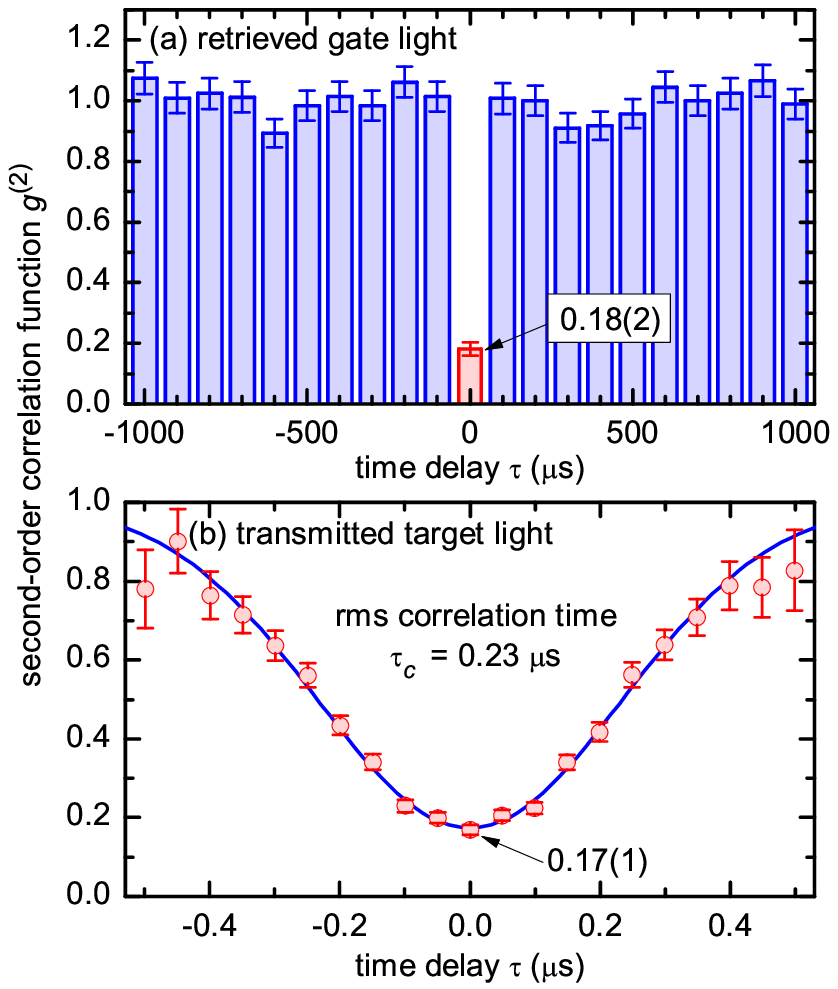}
\caption{\label{fig-sup-g2}
\textbf{Rydberg blockade in correlation functions.} (a) The correlation function $g^{(2)}$ of retrieved photons exhibits a pronounced antibunching minimum due to Rydberg blockade. The incoming pulse contains $N_g= 1.0$ photons on average.
(b) Same for transmitted target light for $N_g=0$ and $N_t=1.7$. The much smaller bin size chosen here reveals the correlation time. The line shows a Gaussian fit.
}
\end{figure}

An independent measurement shows that for $N_t\leq7$ the atomic density (time averaged over all gate-target cycles applied to an atomic sample) decreases as $\rho(N_t)= \rho_0\left(1-{N_t}/{N_1}\right)$ with $N_1= 23$. This decrease is due to the fact that each target signal photon that is absorbed in the atomic gas contributes to photon-recoil heating of the gas. In the shallow dipole trap, evaporation converts this into a loss of atoms. As $OD_b$ is proportional to $\rho$ we expect
\begin{align}
\label{switch-OD-b-Nt}
OD_b(N_t)
= OD_{b,0}\left(1-\frac{N_t}{N_1}\right)
\end{align}
$OD_{b,0}$ is the sum of two contributions. The first contribution is ${2r_b}/{l_a}$ which represents maximal absorption as long as the target photon propagates within the blockade sphere of the stored excitation. The remaining contribution represents the transmission through the rest of the medium. In principle, this contribution should express linear absorption as well as the blockade that the target photons create for each other. However, the smallness of $g^{(2)}(0)$ in Fig.\ \ref{fig-sup-g2}(a) shows that the second contribution comes from a relatively short distance. We therefore neglect the second contribution and obtain the coarse estimate $OD_{b,0}\sim {2r_b}/{l_a} = 5.6$ which is independent of $N_t$.

Combination of Eqs.\ \eqref{switch-OD-b-def} and \eqref{switch-OD-b-Nt} yields the numerator of Eq.\ (1). The denominator in Eq.\ (1) without postselection is shown in Fig.\ \ref{fig-sup-Nt}. It is well modeled by Eq.\ \eqref{propagation-N-out-rapid}. This curve changes only little, if postselection is applied. For the postselected subensemble, we thus obtain
\begin{align}
\label{switch-epsilon-post}
\epsilon^{\text{post}}(N_t)
= \frac{N_t e^{-OD_b(N_t)}}{N_{\text{out}}(N_t)}
\end{align}
with $OD_b(N_t)$ from Eq.\ \eqref{switch-OD-b-Nt} and $N_{\text{out}}(N_t)$ from Eq.\ \eqref{propagation-N-out-rapid}.

The blue line in Fig.\ \ref{fig-Ng-and-Nt}(b) shows a fit of Eq.\ \eqref{switch-epsilon-post} to the postselected subensemble. The curve fits well to the data. The parameters $b= 1.6$, $T_0= 0.30$, and $N_1= 23$ are fixed at the independently determined values. The best-fit value $OD_{b,0}= 5.4$ agrees well with our above estimate. The dominant feature of the fit curve is a moderate deterioration of $\epsilon$ for large $N_t$, which is due to the reduction of the atomic density. This could easily be avoided by averaging over fewer gate-target cycles per atomic sample.

This model is easily extended to describe the total ensemble by
\begin{align}
\label{switch-epsilon}
\epsilon(N_t)
= \frac{(1-p_s)N_{\text{out}}(N_t) + p_s(N_te^{-OD_b(N_t)}+N_0)}
{N_{\text{out}}(N_t)}
\end{align}
with $OD_b(N_t)$ from Eq.\ \eqref{switch-OD-b-Nt} and $N_{\text{out}}(N_t)$ from Eq.\ \eqref{propagation-N-out-rapid}. $p_s(N_g)= 1-p_{s,0}$ is the probability that at least one gate photon was stored. The numerator consists of three terms. The first term represents events for which storage was unsuccessful. In these cases, $N_{\text{trans}}$ is given by $N_{\text{out}}(N_t)$. The second term represents events for which storage was successful. A fair sampling hypothesis yields that for these events $N_{\text{trans}}$ is given by the numerator of Eq.\ \eqref{switch-epsilon-post}. This term has an offset $N_0$ which represents the background photon number during the target time interval for $N_t=0$. $N_0$ is almost completely due to undesired retrieval of stored gate excitations caused by imperfect polarization of the target control light. This term occurs only if storage is successful. We confirmed experimentally that this offset disappears for $N_g=0$. The denominator is the same as the numerator but with $p_s=0$ which follows from $N_g=0$.

We also perform this measurement for $N_t=0$. In this case, the numerator becomes $p_sN_0$ and the measurement yields $p_sN_0= 0.014$ for the total ensemble at $N_g=1.0$. For the postselected subensemble, however, this measurement yields $p_sN_0^{\text{post}}= 0.0028$ which has negligible effect and was therefore not considered in Eq.\ \eqref{switch-epsilon-post}. This difference in the values of $p_sN_0$ is due to the fact that Fig.\ \ref{fig-sup-g2}(a) shows that we almost never store more than one excitation. This excitation can be retrieved only once, either during the target or during the retrieval phase. Hence, events that contribute to the subensemble postselected on retrieval typically do not have an undesired retrieval background during the target phase.

The red line in Fig.\ \ref{fig-Ng-and-Nt}(b) shows a fit of Eq.\ \eqref{switch-epsilon} to the total ensemble. The curve fits well to the data. The parameters $b= 1.6$, $T_0= 0.30$, $N_1= 23$, $OD_b= 5.4$, and $p_sN_0= 0.014$ are fixed. The best-fit value is $p_s= 0.23$.

Assuming that it is unlikely that more than one excitation is stored, we can approximate $N_s\sim p_s$. Combination with $N_g=1.0$ yields an overall storage efficiency $\eta_s= N_s/N_g\sim 0.23$. Combination with Eq.\ \eqref{storage-N-b-rapid} with $b=3.2$, $N_g=1.0$, and $OD_{\text{EIT}}= 0.91$ yields $N_b= 0.56$ and $\eta_{sb}= N_s/N_b\sim 0.41$. The latter is similar to the best-fit value $\eta_{sb}= 0.31$ in Sec.\ \ref{sec-storage}.

Note that our scheme would, in principle, allow for the observation of gain. This is to say that the presence or absence of one gate photon could, in principle, switch the number $N_{\text{trans}}$ of transmitted target signal photons by more than one. A device with this capability offers the possibility of fan-out and is referred to as a single-photon transistor. However, the present performance would require improvements to reach that regime. Presently, the scheme suffers from a nonideal transmission at the EIT resonance, a nonideal storage efficiency, and a fairly short duration of the target pulse. The latter makes it possible to perform postselection, but impossible to transmit a large number of target photons due to self blockade of the target pulse.

\subsection{Rydberg Blockade in Correlation Functions}

\label{sec-correlation}

In addition to the data discussed so far, Rydberg blockade also becomes visible in the normalized second-order correlation function $g^{(2)}(\tau)$ shown in Fig.\ \ref{fig-sup-g2}. Part (a) shows $g^{(2)}(\tau)$ of retrieved gate light, binned over the complete retrieved pulse. Strong antibunching is clearly visible, as also observed in Refs.\ \cite{Dudin:12, Maxwell:13}. The data were taken with a dark time of 0.15 $\mu$s between storage and retrieval. No target light was applied. The polarization of the control light was not switched.

Part (b) shows $g^{(2)}(\tau)$ of transmitted target light with a much smaller bin size. This reveals the correlation time, as also observed in Ref.\ \cite{Peyronel:12}. A Gaussian fit yields an rms correlation time $\tau_c=0.23$ $\mu$s. We compare this with the prediction \cite{Peyronel:12} $\tau_c= 1.05 \sqrt{8\; OD}\; \Gamma/ \Omega_c^2=0.12$ $\mu$s for the parameters of our target pulse. The agreement is fair. The deviation might be due to the inhomogeneity of the medium. Note that $\tau_c$ is quite different from the simple estimate $r_{b,t}/v_g \sim 0.05$ $\mu$s with $v_g=0.3$ km/s from Sec.\ \ref{sec-EIT} for reasons discussed in detail in Ref.\ \cite{Peyronel:12}.

\subsection{Decay of Rydberg Blockade}

\begin{figure}[!t]
\includegraphics[scale=0.95]{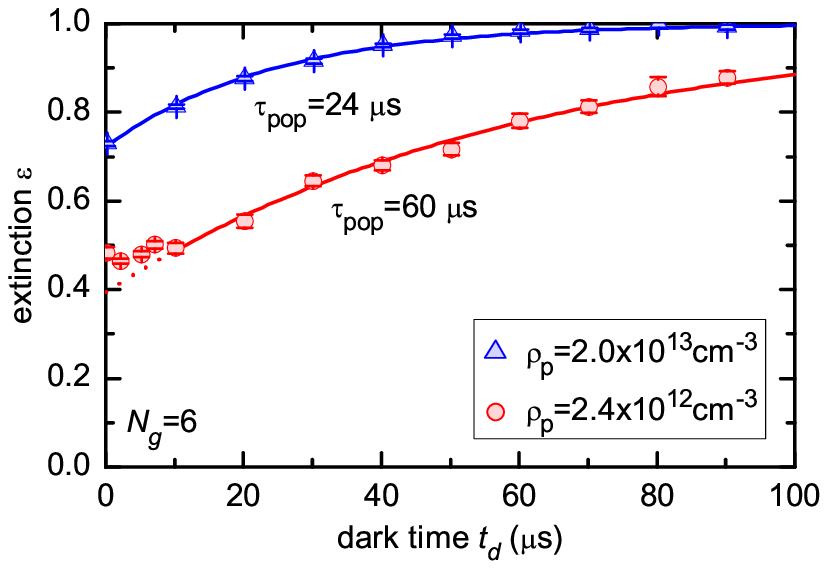}
\caption{\label{fig-sup-population-decay}
\textbf{Decay of Rydberg blockade.}
The extinction $\epsilon$ in the total ensemble decays as a function of the dark time $d_t$ between gate and target pulse. The decay is most likely caused by population decay of Rydberg atoms. Higher density of surrounding ground-state atoms, characterized by the peak density $\rho_p$, causes a faster decay. Exponential fits to the data yield $1/e$ times of $\tau_{\text{pop}}=24$ and 60 $\mu$s. The blue data serve only to illustrate the density dependence. Normally, we do not operate at such high density.
}
\end{figure}

The Rydberg blockade is quite robust. On one hand, Fig.\ \ref{fig-Ng-and-Nt}(b) shows that the extinction $\epsilon$ depends only weakly on $N_t$. On the other hand, Fig.\ \ref{fig-sup-population-decay} shows that it decays only slowly as a function of the dark time $t_d$ between gate and target pulse. This decay is much slower than the dephasing rate observed in Fig.\ \ref{fig-dephasing-rate}. While the decay of retrieval in Fig.\ \ref{fig-dephasing-rate} is sensitive to phase coherence and displays a time scale of $\sim$1 $\mu$s, the decay of blockade in Fig.\ \ref{fig-sup-population-decay} is only sensitive to Rydberg population. An exponential fit (solid line) to the low-density data (red) with $t_d\geq10$ $\mu$s yields a $1/e$ time of 60 $\mu$s. Data for shorter times seem to deviate slightly from the extrapolated fit curve (dotted line).

A model based on spontaneous emission and black-body radiation predicts a 340 $\mu$s lifetime for the $100^2S_{1/2}$ state \cite{Saffman:05}. We obtain a hint at a possible origin of the deviation from the measured decay time of the blockade by comparing this decay time with a measurement at $\sim$8 times higher atomic density (blue data), where the population lifetime is only 24 $\mu$s. This suggests that collisions of a Rydberg atom with ground-state atoms are responsible for the decay. Possible processes are, e.g., associative ionization or ion pair formation \cite{Balewski:13}.

\section{Heralding Probability}

\label{sec-herald}

\begin{figure}[!t]
\includegraphics[scale=0.95]{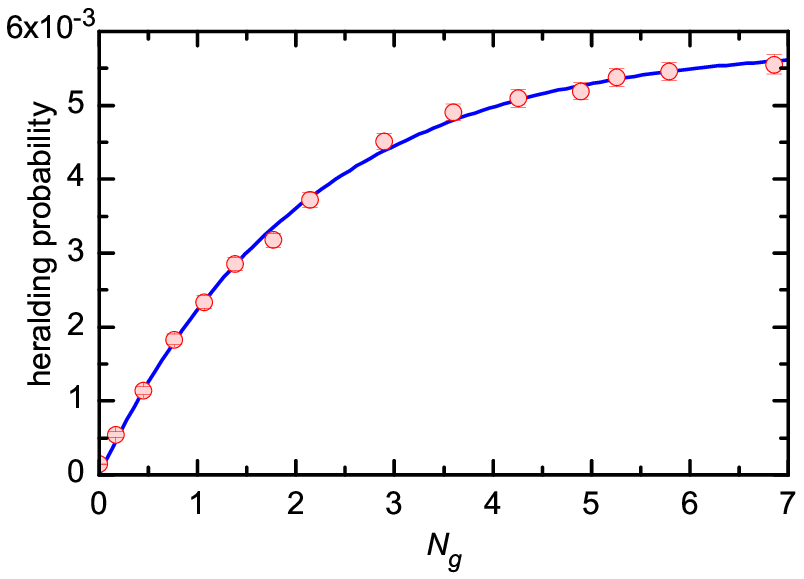}
\caption{\label{fig-sup-herald}
\textbf{Heralding Probability.}
Heralding probability for the data in Fig.\ \ref{fig-Ng-and-Nt}(a). The solid line is a fit of Eq.\ \eqref{herald-ph}.
}
\end{figure}

The heralding probability $p_h$ for the data in Fig.\ \ref{fig-Ng-and-Nt}(a) is shown in Fig.\ \ref{fig-sup-herald}. $p_h$ increases linearly for small $N_g$, whereas it levels off for large $N_g$ due to Rydberg blockade that the gate photons create for each other. This self-blockade of the gate pulse can be modeled by
\begin{align}
\label{herald-ph}
p_h= \eta_{wr} \eta_{\text{det}} N_b
\end{align}
with the write-read efficiency $\eta_{wr}$, the efficiency $\eta_{\text{det}}= 0.27$ for collecting and detecting a transmitted signal photon, and $N_b$ from Eq.\ \eqref{storage-N-b-rapid}. Note that this neglects Rydberg blockade among gate photons after switching the control light back on. This is justified because the average retrieved photon number is low. The solid line shows a corresponding fit with fixed $OD_{\text{EIT}}= 0.91$, yielding best-fit values $b=2.0$ and $\eta_{wr}=0.016$. The value of $b$ obtained here agrees well with the best-fit value obtained from the solid line in Fig.\ \ref{fig-Ng-and-Nt}(a). Various aspects contribute to the value of $\eta_{wr}$. First, the overall storage efficiency is $\eta_s\sim 0.23$, according to Sec.\ \ref{sec-switch}. Second, the write-read efficiency for $t_d\to0$ is $\eta_{wr}= 0.10$ according to the inset in Fig.\ \ref{fig-dephasing-rate}. Third, dephasing in the absence of target control light reduces this to $\eta_{wr}= 0.026$ at $t_d=1.15$ $\mu$s according to the inset in Fig.\ \ref{fig-dephasing-rate}. The best-fit value for $\eta_{wr}$ is slightly worse, mostly due to undesired retrieval of stored excitations caused by imperfect polarization of the target control light.

\section{Choice of Atomic Transitions}

\label{sec-transitions}

The choice of the atomic transitions used in the experiment is based on various considerations that we discuss now. The first issue is related to the choice of the atomic density. On one hand, a good extinction $\epsilon$ for the all-optical switch requires the absorption length $l_{a,t}=1/\varrho \sigma$ during the target pulse to be short compared to the blockade radius $r_{b,t}$. This suggest using large atomic density. On the other hand, Fig.\ \ref{fig-dephasing-rate} shows that a large retrieval signal can only be obtained at low atomic density $\rho$. The density in our experiment is chosen near the point where a further reduction would hardly improve the dephasing rate, as seen in Fig.\ \ref{fig-dephasing-rate}. At this density, we must choose a target signal transition with large absorption cross section $\sigma$ to obtain a sufficiently short absorption length. This is why, we choose $\sigma^-$ polarized target signal light, where the branching ratio is $\xi_t= 1/2$.

To suppress retrieval of a stored gate excitation by the target control light, we switch the control light polarization between gate and target pulse and we use the $P_{1/2}$ state, not the $P_{3/2}$ state. The target signal light couples to the state $|e_t\rangle$ with hyperfine quantum numbers $F=2,m_F=-2$. This stretched state has no component with $m_J=+1/2$, where $J,m_J$ are the fine structure quantum numbers. Hence, $\sigma^-$ polarized control light cannot couple this state to the $100S_{1/2}$ state. Hence, we use $\sigma^+$ polarized control light during the target pulse and the excited state $|e_g\rangle$ for the gate pulse differs from $|e_t\rangle$. As we choose the signal and target beam to have identical propagation directions during gate and target pulse, we choose the $F=2,m_F=0$ state for $|e_g\rangle$. As a result, the absorption cross section on the signal transition is a factor 6 smaller for the gate than for the target pulse. This somewhat reduces the storage efficiency for gate excitations. But, first, that reduction is not very large and, second, this can be improved by postselection.

If, for test purposes, we do not switch the polarization of the control light between gate and target pulse and do not apply target signal light, we observe the undesired retrieval of stored gate excitations. The pulse of outgoing retrieved light has a typical duration of $\sim$0.3 $\mu$s. If we do switch the polarization, the retrieved signal in this time interval is suppressed by a factor of $\sim$7.

\section{Gate-Target Cycles}

\label{sec-gate-target-cycles}

To ensure that our data are taken under similar conditions, we use the same gate-target pulse cycle for the control light in all our measurements. The time for a complete cycle of the control light is always $t_{\text{cyc}}= 100$ $\mu$s. With a 1 $\mu$s decay time for the retrieval and a 60 $\mu$s decay time for the blockade, this suffices to make a cycle independent of previous cycles. In addition, it keeps the value of $\langle V_0\rangle$ in Sec.\ \ref{sec-atoms} negligibly low. In most measurements, we permanently alternate between one cycle with $N_g=0$ and one cycle with $N_g\neq0$. This minimizes the effect of long-term drifts when determining the extinction $\epsilon$.

Fig.\ \ref{fig-sup-population-decay} includes data taken with larger values of $t_d$. To avoid that Rydberg excitations that might unintentionally be stored during a target pulse affect the subsequent gate-target cycle, we use a cycle for signal light that lasts as long as four control-light gate-target cycles. The first control cycle is with signal photon numbers $N_g\neq0$ and $N_t\neq0$, the third control cycle is the reference with $N_g=0$ and $N_t\neq0$. The remaining second and fourth cycle are with $N_g=N_t=0$ which makes sure that the time-averaged control intensity remains the same as in the other measurement but sufficient time elapses between application of signal light that all Rydberg excitations can decay.

The value of $\epsilon$ quoted in the caption of Fig.\ \ref{fig-time-traces} is determined by processing data from 1841 atomic samples. We produce approximately 3 atomic samples per minute, so that this corresponds to approximately 10 hours of data acquisition. For each atomic sample, we process data between 50 and 950 ms after starting the gate-target cycles. With $t_{\text{cyc}}= 100$ $\mu$s, we process 9000 cycles per atomic sample; one half of them with $N_g=1$ and the other half with $N_g=0$. The average number of detected target photons in one gate-target cycle is $\sim$0.1, so that we cannot calculate a value for $\epsilon$ for each cycle directly because the denominator is zero too often. We therefore sum up all detector clicks with $N_g=1$ from one atomic sample and do the same for all detector clicks with $N_g=0$ from the same atomic sample. Dividing the first number by the second yields one value of $\epsilon$ for each atomic sample. For this sample of 1841 values of $\epsilon$, we calculate the sample mean and its standard error. For the postselected data, the count rates are even lower and we add the heralded events from all atomic samples before performing the division to calculate $\epsilon$. The error bar for the postselected data is based on the assumption of shot noise on the number of contributing events.

\end{document}